\documentclass {llncs}
\usepackage{amsfonts,latexsym, amssymb}
\usepackage{proof}

\def\FV#1{FV({#1})}           
\def\TV#1{TV({#1})}           
\def\no#1#2{no({#1, #2})}     
\def\dep#1#2{d(#1, #2)}       
\def\depth#1{d(#1)}           
\def\rank#1{rank({#1})}       
\def\WE#1#2{W(#1, #2)}        
\def\taille#1{|#1|}           

\newcommand\Nat{N}            

\def\MES#1#2{m(#1, #2)}       
\def\ME#1{M(#1,n)}              
\def\nlet#1{nlet(#1)}         

\newcommand{\ma}{\mathcal}

\def\size#1{size({#1})}

\def\vect#1{\overrightarrow{#1}}      

\def\lt#1#2#3{\mbox{let} \, #1 \, \mbox{be} \, #2 \, \mbox{in} \, #3}        
\newcommand\lef{\mbox{ left }}
\newcommand\righ{\mbox{ right }}

\def\lta#1{\mbox{let} \, #1 \, \mbox{be} \, }
\def\casleft#1#2{\mbox{ left } \, #1 \, \mbox{in} \, #2 \, }
\def\casright#1#2{\mbox{ right } \, #1 \, \mbox{in} \, #2 \, }   

\newcommand\red{\; \rightarrow \;}                                           
\newcommand\redun{\; \rightarrow ^{1} \;}                                 
\newcommand\redstar{\; \rightarrow ^{\star} \;}
\def\sub#1#2 {[ #1 \slash #2 ]}                                            

\def\size#1{|#1|} 

\newcommand {\fm} {\multimap}

\newcommand{\te} {\otimes}

\newcommand{\av} {\mathbin{\&}}
\newcommand{\pl} {\mathbin{\oplus}}
\newcommand{\bs} {\mathord{!}}

\newcommand{\al} {\alpha}
\newcommand{\la} {\lambda}

\newtheorem{theo}{Theorem}
\newtheorem{de}{Definition}
\newtheorem{prop}[theo]{Proposition}
\newtheorem{lem}[theo]{Lemma}
\newtheorem{Cor}[theo]{Corollary}

\title{ Soft lambda-calculus: a language for\\ polynomial time computation}
\author{Patrick Baillot \and Virgile Mogbil}
\author{Patrick Baillot \and Virgile Mogbil}
\institute{Laboratoire d'Informatique de Paris-Nord UMR 7030 CNRS\\
Universit\'e Paris XIII - Institut Galil\'ee,
99  Avenue Jean-Baptiste Cl\'ement,\\
93430 Villetaneuse, France.
\email{\{patrick.baillot,virgile.mogbil\}@lipn.univ-paris13.fr}
\thanks{Work partially supported by Action Sp\'ecifique CNRS \textit{M\'ethodes formelles pour la Mobilit\'e} and ACI 
S\'ecurit\'e Informatique CRISS.}
 }
\begin{document}
\bibliographystyle{alpha}
\maketitle

\begin{abstract}
 Soft linear logic ([Lafont02]) is a subsystem of linear logic
characterizing the class PTIME. We introduce \emph{soft
lambda-calculus} as a calculus typable in the intuitionistic and
affine variant of this logic.  We prove that the (untyped) terms of
this calculus are reducible in polynomial time. We then extend the
type system of Soft logic with recursive types. This allows us to
consider non-standard types for representing lists.  Using these
datatypes we examine the concrete expressivity of Soft lambda-calculus
with the example of the insertion sort algorithm.
\end{abstract}

\section{Introduction}\label{introduction}
 With the advent of global computing there are an increasing variety
 of situations where one would need to be able to obtain formal bounds
 on resource usage by programs: for instance before running code
 originating from untrusted source or in settings where memory or time
 is constrained, like in embedded systems or synchronous systems.

Some cornerstones for this goal have been laid by the work on
\textit{Implicit Computational Complexity} (ICC) as carried out by
several authors since the 1990s (\cite{Leivant94},
\cite{LeivantMarion93}, \cite{BellantoniCook} among others). This
field aims at studying languages and calculi in which all programs
fall into a given complexity class. The most studied case has
naturally been that of deterministic polynomial time complexity (PTIME
class). We can in particular distinguish two important lines of
work. The first one deals with primitive recursion and proposes
restrictions on primitive recursion such that the functions definable
are those of PTIME: this is the approach of Bellantoni-Cook
(\cite{BellantoniCook}) and subsequent extensions
(\cite{Hofmann00},\cite{BNS00}).

Another line is that of Linear logic (LL)(\cite{Girard87a}). By the
 Curry-Howard correspondence proofs in this logic can be seen as
 programs. Linear logic provides a way of controlling duplication of
 arguments thanks to specific modalities (called
 \textit{exponentials}). It is possible to consider variants of LL
 with alternative, stricter rules for modalities, for which all
 proofs-programs can be run in polynomial time. 
Light linear logic, introduced by Girard (\cite{Girard95a}) is one of
these systems. It was later simplified by Asperti into Light affine
logic (\cite{AspertiRoversi02},\cite{Asperti98}) which allows full weakening
(that is to say erasing of arguments). However formulas in this system
are quite complicated as there are two modalities, instead of just
one in intuitionistic linear logic. More recently Lafont introduced
Soft linear logic (SLL) (\cite{Lafont02}), a simpler system which uses
the same language of formulas as Linear logic and is polytime. It can
in fact be seen as a subsystem of linear logic or of Bounded linear
logic (\cite{GirardScedrovScott}).

In all these approaches it is shown that the terms of the calculus can be
evaluated in polynomial time. A completeness result is then proved by
simulating in the calculus a standard model for PTIME computation such as PTIME
Turing machines. It follows that all PTIME \textit{functions} are representable in
 the calculus, which establishes its expressivity.

 However if this completeness argument is convincing for
characterization of complexity classes of functions, it is rather
unsatisfactory when we are interested in the use of Implicit
Computational Complexity for the study of program properties. Indeed
it is not so appealing to program in a new language via the encoding
of Turing machines \dots One would prefer to be able to take advantage
of the features of the language: for the variants of Linear logic for
instance we have at hand abstract datatypes and structural recursion,
higher-order and polymorphism.

Some authors have observed that common algorithms such as insertion
sort or quicksort are not directly representable in the
Bellantoni-Cook approach (see for instance
\cite{Hofmann99}). Important contributions to the study of programming
aspects of Implicit computational complexity have been done in
particular by Jones (\cite{Jones97}), Hofmann (\cite{Hofmann99}) and
Marion (\cite{Marionhabilitation}). For instance Hofmann proposed
languages using linear type systems with a specific type for space
unit, which enabled him to characterize non-size increasing
computation with various time complexity bounds. This approach allows
to represent several standard algorithms.

Here we are interested in investigating the programming possibilities
 offered by Soft linear logic. In \cite{Lafont02} this system is
 defined with sequent-calculus and the results are proved using
 proof-nets, a graph representation of proofs. In order to make the
 study of programming easier we propose a lambda-calculus
 presentation. We extend for that usual lambda-calculus with new
 constructs corresponding to the exponential rules of SLL.  The
 resulting calculus is called Soft lambda-calculus and can be typed in
 SLL. Actually we choose here the affine variant of Soft logic as it
 is more flexible and has the same properties.  Our Soft
 lambda-calculus is inspired from Terui's Light affine lambda-calculus
 (\cite{Terui01}), which is a calculus with a polynomial bound on
 reduction sequences that can be typed in Light affine logic.
 
 \textbf{ Outline}. In section \ref{softlambda-calculus} we define
 soft lambda-calculus and its type-assignment system. Then in section
 \ref{bounds} we prove that the length of any reduction sequence of a
 term is bounded by a polynomial applied to the size of the term. In
 section \ref{extension} we extend the type system and add recursive
 typing. Finally in section \ref{datatypes} we examine datatypes for
 lists and propose a new datatype with which we program the insertion
 sort.

\textbf{Acknowledgements}. We wish to thank Marcel Masseron for the
stimulating discussions we had together on Soft linear logic and which
led to the present paper. Thanks also to Kazushige Terui for his
useful comments and suggestions.

\section{Soft lambda-calculus}\label{softlambda-calculus}  

The introduction of our calculus will done be in two steps (as in
\cite{Terui01}): first we will define a grammar of pseudo-terms and
then we will distinguish terms among pseudo-terms.

 The \emph{pseudo-terms} are defined by the grammar:
$$t, t' ::= x \, | \, \lambda x \, t \, | \, (t \; t') \,|\, \bs t \,|\, \lt{t} {\bs x} {t'} $$

 For a pseudo-term $t$ we consider:
\begin{itemize}
\item its set of free variables $\FV{t}$;
\item for a variable $x$ the number of free occurrences $\no{x}{t}$ of $x$ in $t$. 
\end{itemize}
In the pseudo-term $\lt{u} {\bs x} {t_1}$, the variable $x$ is bound:
$$\FV{\lt{u} {\bs x} {t_1}}= \FV{u} \cup \FV{t_1} \backslash \{x\}$$ 

If $t$ is of the form $\lt{u}{x}{t_1}$ we say that $t$ is a \textit{let expression}.

If $\vect{t}$ and $\vect{x}$ respectively denote finite sequences of
same length $(t_1, \dots, t_n)$ and $(x_1, \dots, x_n)$, then
$\lt{\vect{t}} {\bs \vect{x}} {t'}$ will be an abbreviation for $n$
consecutive $let$ expressions on $t_i$s and $x_i$s: $\lt{t_1} {x_1} {\lt{t_2} {x_2} {\dots} }{t'}$ .
 
We define the \emph{size} $\taille{t}$ of a pseudo-term $t$ by:

\begin{center}
$\begin{array}{ll}
\taille{x}=1  & \taille{\lambda x \, t}=\taille{t}+1 \\
 \taille{(t \; t')}=\taille{t}+\taille{t'} & \taille{\bs t}=\taille{t}+1\\
\taille{\lt{t} {\bs x} {t'}}=\taille{t}+\taille{t'}+1 & 
\end{array}$
\end{center}

We will type these pseudo-terms in intuitionistic
soft \emph{affine} logic (ISAL).  The formulas are given by the
following grammar:
$$
T::= \al \; |\;  T \fm T \; |\;  \forall \al. T \; |\;  \bs \; T $$
We
choose the affine variant of Soft linear logic, which means permitting full
weakening, to allow for more programming facility. This does not
change the polytime nature of the system, as was already the case for
light logic (\cite{Asperti98,Terui01}).

We give the typing rules in a sequent calculus presentation. It offers the advantage
of being closer to the logic. It is not so convenient for type-inference, but it is not
our purpose in this paper. The typing rules are given on Figure \ref{ISALrules}.
 
 \begin{figure}[ht]
  \begin{center}
\fbox{
\begin{tabular}{l@{}ll}
  &{\infer[\mbox{(variable)}]{x:A \vdash x:A}{}} &
  {\infer[\mbox{(Cut)}]{\Gamma,\Delta \vdash u[t/x]:B}
  {\Gamma \vdash t:A & \Delta, x:A \vdash u:B}}\\
 &&\\
 &{\infer[\mbox{(right arrow)}]{\Gamma \vdash \la x. t: A \fm B }
 {\Gamma, x:A \vdash t:B}}
  & {\infer[\mbox{(left arrow)}]{\Gamma, \Delta, y: A \fm B \vdash t\sub{(yu)}{x} :C }
  {\Gamma , x:B \vdash t:C & \Delta \vdash u:A}}\\
&{\infer[\mbox{(weak.)}]{\Gamma, x:A\vdash t: B }
 {\Gamma \vdash t:B}}
  & \\[1ex]
&{\infer[\mbox{(mplex)}]{y:\bs{A}, \Gamma \vdash \lt{y} {\bs x} {t\sub{x}{x_1,\dots ,x_n} }:B }{x_1:A, \dots, x_n:A ,\Gamma \vdash t:B }} 
 & {\infer[\mbox{( prom.)}]{y_1: \bs{A_1}, \dots, y_n:
\bs{A_n} \vdash  \lt{\vect{y}} {\bs \vect{x}} {t}: \bs{B}}{x_1: A_1, \dots, x_n: A_n \vdash t:
B}}\\
&{\infer[\mbox{left $\forall$}]{x:\forall \alpha .A , \Gamma \vdash t:B }{x:A, \Gamma \vdash t:B }} 
 & {\infer[\mbox{right $\forall$ (*)}]{\Gamma \vdash  t:\forall \alpha. B}{\Gamma \vdash t:B}}
\end{tabular}
}
\end{center}
  \caption{ISAL typing rules}\label{ISALrules}
\vspace*{-3mm}
\end{figure}

For (right $\forall$) we have the condition: 
\begin{center}(*) $\alpha$ does not appear free in $\Gamma$. \end{center}

Observe that the $let$ expression is used to interpret both the
\emph{multiplexing} (mplex) and the \emph{promotion} (prom.) logical
rules. We could distinguish two different kinds of \textit{let} but we prefer
to have a small calculus.

For instance one can consider for unary integers the usual type of Linear logic:
 \begin{center}$\Nat=\forall \alpha. \bs(\alpha \fm \alpha)\fm \alpha \fm \alpha$\end{center}
 The integer $n$ is represented by the following pseudo-term of type
 $\Nat$, with $n$ occurrences of $s'$:
\begin{center}$\la s. \la x. \lt{s}{\bs s'}{(s' \; (s' \;(s' \dots x)\dots )}$\end{center}
Among pseudo-terms we define a subclass of \emph{terms}. These will be
defined inductively together with a notion of \emph{temporary
  variables}.  The temporary variables of a term $t$, $\TV{t}$, will
be part of the free variables of $t$: $\TV{t} \subseteq \FV{t}$.

\begin{de}          
 The set $\mathcal{T}$ of terms is the smallest subset of pseudo-terms such that:
\begin{itemize}
\item $x \in \mathcal{T}; \; \mbox{then } \TV{x}=\emptyset;$ 
\item $\la x. t \in \mathcal{T} $ iff: $x \notin \TV{t}$, $t \in
\mathcal{T} $ and $\no{x}{t} \leq 1$;

then $\TV{\la x.t}=\TV{t};$
\item  $ (t_1 \; t_2) \in \mathcal{T}$ iff:   $t_1, t_2 \in \mathcal{T} $, $\TV{t_1} \cap \FV{t_2}=\emptyset$,
$\FV{t_1} \cap \TV{t_2}=\emptyset$;

 then $\TV{(t_1 \; t_2)}=\TV{t_1}\cup \TV{t_2}$;
\item  $ \bs t \in \mathcal{T}$ iff: $t \in \mathcal{T} $, $\TV{t}=\emptyset \mbox{ and } \forall x \in \FV{t}, \no{x}{t}=1$;

then $\TV{\bs t}=\FV{t}$;

\item $\lt{t_1} {\bs x} {t_2} \in \mathcal{T}$ iff: $t_1, t_2 \in \mathcal{T} $, $\TV{t_1} \cap \FV{t_2}=\emptyset$,
$\FV{t_1} \cap \TV{t_2}=\emptyset$;       
 
 then $\TV{\lt{t_1} {\bs x} {t_2}}=\TV{t_1}\cup (\TV{t_2} \backslash \{x\})$.
\end{itemize} 
\end{de} 
Basically the ideas behind the definition of terms are that:
\begin{itemize}
\item one can abstract only on a variable that is not temporary and which has at most one occurrence,
\item one can apply $\bs$ to a term which has no temporary variable and whose free variables
have at most one occurrence; the variables then become temporary;
\item the only way to get rid of a temporary variable is to bind it using a \textit{let}
expression.
\end{itemize}
 It follows from the definition that temporary variables in a term are linear:
\begin{lem}\label{tempvariableslinear}
 If $t$ is a term and $x \in \TV{t}$, then $\no{x}{t}=1$.
\end{lem}
The definition of depth will be useful later when discussing reduction:
\begin{de}\label{defdepth}
 Let $t$ be a term and $u$ be an \emph{occurrence} of subterm of $t$. We call \emph{depth} of $u$ in $t$, $\dep{u}{t}$
the number $d$ of subterms $v$ of $t$ such that $u$ is a subterm of $v$ and $v$ is of the form $\bs v'$. 

The depth $\depth{t}$ of a term $t$ is the maximum of $\dep{u}{t}$ for $u$ subterms of $t$.
\end{de}

 For instance: for $t= \bs (\la f. \la x. \lt{f}{\bs f'}{\bs(f' x)}$ and $u=(f' x)$, we have $\dep{u}{t} = 2$.

We can then observe that:
\begin{prop}\label{depthvariables}
 Let $t$ be a term. If $x$ belongs to $\FV{t}$ and $x_0$ denotes an occurrence of $x$ in $t$, then $\dep{x_0}{t}\leq 1$.

Moreover all occurrences of $x$ in $t$ have the same depth, that we can therefore denote by $\dep{x}{t}$, and we have: $\dep{x}{t}=1$ iff $x \in \TV{t}$.
\end{prop}
In fact we will focus our attention on specific terms:
\begin{de}\label{defwellformed}
  A term $t$ is \emph{well-formed} if we have: 
$$\TV{t}=\emptyset \mbox { and } \forall x \in \FV{t}, \; \no{x}{t}=1.$$
\end{de}
Note that to transform an arbitrary term into a well-formed one, one
only needs to add enough \emph{let} expressions.

We have the following properties on terms and substitution:
\begin{lem}\label{lemmasubtermbang}
 If $t$ is a term and $t=\bs t_1$, then $t_1$ is a well-formed term.
\end{lem}

\begin{lem}\label{substitution1}
 If we have:
\begin{itemize}
\item $t$, $u$ terms,
\item $\TV{u}=\emptyset$,
\item $x \notin \TV{t}$, 
\item $\FV{u} \cap \TV{t}=\emptyset$,        
\end{itemize}
 then: $t\sub{u}{x} $ is a term and $\TV{t \sub{u}{x} }=\TV{t}$.
\end{lem}

We can then check the following:

\begin{prop}\label{typedimplieswellformed}
 If $t$ is a pseudo-term such that in ISAL we have $\Gamma \vdash t:A$, then $t$ is a well-formed term.
\end{prop}
\begin{proof}
 by induction on the type derivation, using the definition of terms and for the case of the (cut) 
and (leftarrow) rules the
lemma \ref{substitution1}.
\end{proof}

We will also need in the sequel two variants of lemma \ref{substitution1}:

\begin{lem}\label{substitution1bis}     
 If we have:
\begin{itemize}
\item $t$, $u$ terms,
\item $x \notin \TV{t}$, 
\item $\no{x}{t}=1$,
\item $\FV{u} \cap \TV{t}=\emptyset$,        
\item $\TV{u} \cap \FV{t}= \emptyset$,
\end{itemize}
 then: $t\sub{u}{x} $ is a term and $\TV{t \sub{u}{x} }=\TV{t}\cup \TV{u}$.
\end{lem}
 Note that the main difference with lemma \ref{substitution1} is that we have here the assumption $\no{x}{t}=1$.
\begin{lem}\label{substitution2}
 If we have:
\begin{itemize}
\item $t$ is a term and $u$ is a well-formed term,
\item $\FV{t}\cap \FV{u}=\emptyset$,          
\item $x \in \TV{t}$ 
\end{itemize}
 then: $t\sub{u}{x} $ is a term and $\TV{t \sub{u}{x} }=\TV{t}\backslash \{x\}\cup \FV{u}$.
\end{lem}

We now consider the contextual one-step reduction relation $\redun$ defined on
pseudo-terms by the rules of figure \ref{reductionrules}. The rules
(com1) and (com2) are the commutation rules.  The relation $\red$ is
the transitive closure of $\redun$.

\begin{figure}[ht]         
  \begin{center}
\fbox{\begin{tabular}{ll@{}l}
 ($\beta$): &\quad & $((\la x.t) \; u)  \redun t\sub{u}{x} $\\
 (bang) :   &\quad &  $\lt{\bs u} {\bs x} {t} \redun t\sub{u}{x} $\\
(com1):     &      &$\lt{(\lt{t_1}{\bs y}{t_2})}{ \bs x}{t_3} \redun \lt{t_1}{\bs y}{(\lt{t_2}{\bs x}{t_3})}$ \\
 (com2):    &      & $(\lt{t_1}{\bs x}{t_2})t_3 \redun \lt{t_1}{\bs x}{(t_2 \; t_3)}$ 
\end{tabular}
}
\end{center}
 \caption{reduction rules}\label{reductionrules}
\vspace*{-3mm}
\end{figure}


We have:
\begin{lem}\label{termspreservedbyreduction}
The reduction is well defined on terms (the result of a reduction step on a term is a term).
 Furthermore, if $t$ is a well-formed term and $t \redun t'$, then $t'$ is well-formed.
\end{lem}

Finally we have:
\begin{prop}[local confluence]\label{localconfluence}
The reduction relation $\redun$ on terms is locally confluent:
if $t \redun t'_1$ and $t \redun t'_2$ then there exists $t'$ such that
$t'_1 \red t'$ and $t'_2 \red t'$.
\end{prop}

\section{Bounds on the reduction}\label{bounds}

We want to find a polynomial bound on the length of reduction sequences
of terms, similar to that holding for SLL proof-nets (\cite{Lafont02}).
For that we
must define a parameter on terms corresponding to the arity of the
multiplexing links in SLL proof-nets.
\begin{de}
 The \emph{rank} $\rank{t}$ of a term $t$ is defined inductively by:


$\begin{array}{lll}
  \rank{x} & = \; & 0 \\
 \rank{\la x.t}   &=  & \rank{t}\\
\rank{(t_1 t_2)}  &=& \max(\rank{t_1}, \rank{t_2}) \\
\rank{\bs t} &=& \rank{t}\\
\rank{\lt{u}{\bs x}{t_1}}&=&
\left\lbrace\begin{array}{ll}
 \max(\rank{u}, \rank{t_1}) & \mbox{ if } x\in\TV{t_1} \\
 \max(\rank{u}, \rank{t_1}, \no{x}{t_1}) &\mbox{ if } x\notin\TV{t_1}
             \end{array}\right.

\end{array}
$
\end{de}

The first case in the definition of $\rank{\lt{u}{\bs x}{t_1}}$
corresponds to a promotion, while the second one corresponds to a
multiplexing and is the key case in this definition.

 To establish the bound we will adapt the argument given by Lafont for proof-nets. First we define for a term $t$ and an
integer $n$ the \emph{weight} $\WE{t}{n}$ by:

$$\begin{array}{lll}
  \WE{x}{n} & = \; & 1 \\
 \WE{\la x.t}{n}   &=  & \WE{t}{n}+1 \\
\WE{\bs u}{n}  &=& n\WE{u}{n}+1 \\
\WE{(t_1 t_2)}{n} &=& \WE{t_1}{n}+\WE{t_2}{n}\\
\WE{\lt{u}{\bs x}{t_1}}{n}&=& \WE{u}{n}+\WE{t_1}{n} \\
\end{array}
$$

 We have the following key lemma:
\begin{lem}\label{lemKey}
 Let $t$ be a term and $n\geqslant\rank{t}$.
\begin{enumerate}
\item if $x \notin \TV{t}$ and $\no{x}{t}=k$, then:
$$ \WE{t\sub{u}{x} }{n}\leqslant\WE{t}{n} +k\WE{u}{n}$$
\item if $x \in \TV{t}$ then:
$$ \WE{t\sub{u}{x} }{n}\leqslant\WE{t}{n} +n\WE{u}{n}$$
\end{enumerate}
\end{lem}

We give the proof of this lemma in Appendix \ref{prooflemKey}.

\begin{prop}\label{propWeightDecrease}
 Let $t$ be a term and $n\geqslant\rank{t}$. If $t\redun t'$ by a $(\beta )$ or
 $(bang)$ reduction rule then $\WE{t'}{n} <\WE{t}{n}$.
\end{prop}

\begin{proof}
  If $t\stackrel{\sigma}{\red} t'$ with $\sigma=(\beta)$ or $(bang)$
  then let $r$ denote the redex reduced inside $t$. The form of $t$ is
  \mbox{$t_0\sub{r}{y} $} with $\no{y}{t_0}=1$ and $t'=t_0\sub{r'}{y}
  $ where $r\stackrel{\sigma}{\red} r'$.

The result is obtained by induction on the term $t_0$ for a given
$n\geqslant\rank{t}$:

let us consider the basic case $t_0=y$, i.e. $t=r$ using the definitions of
terms and rank, and lemma \ref{lemKey}:

 for instance for a (bang) reduction rule, 

$\begin{array}{lll}
r &=& \lt{\bs u}{\bs x}{r_1} \\
r' &=& r_1\sub{u}{x} \\
\WE{r}{n} &=& \WE{\lt{\bs u}{\bs x}{r_1}}{n} =n.\WE{u}{n} +1+\WE{r_1}{n}
\end{array}$. 

If $x\in\TV{r_1}$ then by lemma \ref{lemKey} $\WE{r'}{n} <\WE{r}{n}$, 
else $x\in\FV{r_1}\backslash\TV{r_1}$ and 

$\begin{array}{lll}
\WE{r'}{n} &\leqslant &\WE{r_1}{n} +\no{x}{r_1} .\WE{u}{n}\\
           &\leqslant &\WE{r_1}{n} +\rank{r}.\WE{u}{n}\\
           &\leqslant & \WE{r_1}{n} +n.\WE{u}{n} \\
           &< &\WE{r}{n}
\end{array}$

In the non basic cases, i.e $t_0\not =y$, we can remark that 
 $\WE{t_0\sub{r}{x} }{n}$ is a strictly increasing function 
of $\WE{r}{n}$. For instance:

if $t_0=(y \; t_1)$ then
$\WE{t'}{n} =\WE{t_0\sub{r'}{y} }{n} =\WE{(r't_1}{n} =\WE{r'}{n} +\WE{t_1}{n}
<\WE{r}{n} +\WE{t_1}{n}$ i.e. $\WE{t'}{n} <\WE{t}{n}$.
\end{proof}

For the commutation rules we have $\WE{t'}{n} =\WE{t}{n}$. 
So we need to use a measure of the commutations in a reduction sequence to
be able to bound the global length. We
make an adaptation of the weight used in \cite{Terui01}. 

Given an integer $n$ and  a term $t$, for each subterm occurrence in $t$ of the form
$t_1\equiv\lt{u}{\bs x}{t_2}$, we define the \emph{measure} of $t_1$ in
$t$ by: 

$$\MES{t_1}{t} =\WE{t}{n} -\WE{t_2}{n}$$
and $\ME{t}$ the \emph{measure} of $t$ by the sum of $\MES{t_1}{t}$ for all subterms $t_1$ of
$t$ which are \textit{let} expressions.

\begin{prop}\label{propMesureDecrease}
  Let $t$ be a term and $n\geqslant\rank{t}$. If $t\redun t'$ by a
  commutation reduction rule then $\ME{t'} <\ME{t}$.
\end{prop}


Given a term $t$ we denote by $\nlet{t}$ the number of subterm
occurrences of \textit{let} expressions in $t$.

\begin{lem}\label{lemBoundNlet}
Let $t$ be a term and $n\geqslant 1$. We have
$\nlet{t}\leqslant\WE{t}{n} -1$.
\end{lem}

\begin{prop}\label{propBoundWeight}
 If $t$ is a term and $p=\depth{t}$, $k=\WE{t}{1} $, and $n\geqslant 1$ then:
$$ \WE{t}{n}\leqslant k. n^p $$
\end{prop}

\begin{proof}
Let $n\geqslant 1$. By induction on the term, using definitions of weight
and depth: if $t=!t_1$ then

 $\begin{array}{lll}
\WE{t}{n}=n.\WE{t_1}{n}+1 &\stackrel{i.h.}{\leqslant}&
\WE{t_1}{1} .n^{d(t_1)+1}+1 \quad \mbox{because $n\geqslant 1$}\\
 &\leqslant& (\WE{t_1}{1} +1).n^p\\
&\leqslant & \WE{t}{1}.n^p
\end{array}$

  The other cases are immediate.
\end{proof}

\begin{theo}\label{theopolytime}[Polytime reduction]

 For any integer $d$ there is a polynomial $P_d$ (with degree linear in $d$) such that:
 
for any term $t$ of depth $d$, any sequence of reductions of $t$ has length bounded by $P_d(\taille{t})$.
\end{theo}

\begin{proof}
  Let $t$ be a term of depth $d$ and $n\geqslant\rank{t}$. We will
  call \textit{round} a sequence of reductions and \textit{proper
    round} a non empty reductions sequence of $(\beta)$ and $(bang)$
  reductions.
  
  If $t\stackrel{\sigma}{\red }t'$ then there is an integer $l$ such
  that $\sigma$ can be described by an alternate sequence of
  commutation rules rounds and proper rounds as follows:
$$t=t_1\stackrel{(com)}{\redstar} t_2
\stackrel{(\beta),(!)}{\redstar} t_3\ \ldots\  
t_{2i+1}\stackrel{(com)}{\redstar} t_{2i+2}
\stackrel{(\beta),(!)}{\redstar} t_{2i+3}\ \ldots\ 
t_{2l+1}\stackrel{(com)}{\redstar} t_{2l+2}=t'$$

Remark that the alternate sequence starts and finishes with a
commutation rules round. The sequence $\sigma$ contains $l$ 
proper rounds. Because each such round strictly decreases the weight
of $t$ (Prop.\ref{propWeightDecrease}) and the commutation rules leave
the weight unchanged we have $l\leqslant\WE{t}{n}$. Moreover 
the length of all proper rounds in $\sigma$ is bounded by $\WE{t}{n}$.

On the other hand we have by definition and lemma \ref{lemBoundNlet}:
$$\ME{t'} <\nlet{t'} .\WE{t'}{n} <(\WE{t'}{n} )^2 -\WE{t'}{n} 
<(\WE{t}{n} )^2 -\WE{t}{n} .$$

There are at most $(l+1)$ commutation rules rounds, so by Prop. 
\ref{propMesureDecrease}  the length of all such
rounds is bounded by $(l+1).((\WE{t}{n} )^2 -\WE{t}{n} )$. Then we have 
$$\size{\sigma}\leqslant (l+1).((\WE{t}{n } )^2 -\WE{t}{n} )+\WE{t}{n}
\leqslant (\WE{t}{n} )^3$$
Finally this result can be applied to any $n\geqslant\rank{t}$. 
Consider $n=\size{t}$, by prop.\ref{propBoundWeight} we obtain that
$$\size{\sigma}\leqslant (\WE{t}{1} )^3 .(\size{t} )^{3d}
\leqslant (\size{t} )^{3(d+1)}$$

where $d=d(t)$.
\end{proof}

\begin{remark}
If a term $t$ of depth $d$ corresponds to a program and $u$ to an argument
 such that $\depth{u}\leqslant \depth{t}$, then $(t\; u)$ normalizes in at most 
$Q_d(\size{u})$ steps for some polynomial $Q_d$: 

by the previous theorem if $(t\; u)\stackrel{\sigma}{\red} t'$ then 
$\size{\sigma}\leqslant (\size{t} +\size{u} )^{3(d+1)}$ because 
$\depth{(t\; u)} =\depth{t} =d$. 
Let $Q_d(X)$ be the following polynomial : 

 \begin{center}$Q_d(X)=(X+\size{t} )^{3(d+1)}$.\end{center}
\end{remark}

Note that theorem \ref{theopolytime} shows that the calculus is
 \textit{strongly} polytime in the sense of \cite{Terui01}: there exists a polynomial
bounding the length of \textit{any} reduction sequence (no matter the reduction strategy).
 An obvious consequence is then:  

\begin{Cor}[Strong normalization]
The terms of soft lambda calculus are strongly normalizing.
\end{Cor}

\begin{Cor}[Confluence property]
If a term $t$ is such that $t\red u$ and $t\red v$ then there exists a term $w$ such
that $u\red w$ and $v\red w$.
\end{Cor}

\begin{proof}
By local confluence (Proposition \ref{localconfluence}) and strong normalization.
\end{proof}

\section{Extension of the calculus}\label{extension}

Thanks to full weakening, the connectives $\te$, $\av$, $\pl$,
$\exists$ and the constant $1$ are definable from $\{\fm, \forall \}$
(\cite{Asperti98}, \cite{TeruiPhd}):
\begin{eqnarray*}
 \exists \beta.A & =& \forall \alpha.(\forall \beta. (A \fm \alpha)\fm \alpha) \\
 A \te B & =& \forall \alpha. ((A \fm B \fm \alpha) \fm \alpha) \\
   1     & =& \forall \al. (\al \fm \al) \\
 A \pl B&=& \forall \alpha.((A \fm \alpha) \fm (B \fm \alpha) \fm \alpha)\\
 A \av B & =& \exists \alpha.((\alpha \fm A) \te (\alpha \fm B)\te \alpha) 
\end{eqnarray*} 

 We use as syntactic sugar the following new constructions on terms:

$\begin{array}{llll}
 t_1 \te t_2 ,    \qquad &\mbox{let }u \mbox{ be } &{x_1 \te x_2}\mbox{ in }{t} , \\
 \lef{t},  \qquad \qquad  & \lta{u} &\casleft{x}{t_1}\\
 \righ{t} ,                           &         &  \casright{y}{t_2};\\
\end{array}
$


We then have the new typing rules of figure \ref{derivedrules}.

\begin{figure}[ht]
  \begin{center}
\fbox{
\begin{tabular}{l@{}ll}
  &{\infer[\mbox{(left $\te$)}]{\Gamma, x: A_1 \te A_2 \vdash \lt{x}{x_1 \te x_2}{t}: B }
 {\Gamma, x_1:A_1, x_2:A_2 \vdash t:B}}
  & {\infer[\mbox{(right $\te$)}]{\Gamma_1, \Gamma_2 \vdash t_1 \te t_2:A_1 \te A_2 }
  {\Gamma_1 \vdash t_1:A_1 & \Gamma_2 \vdash t_2:A_2}}\\
&&\\
&{\infer[\mbox{(left $\pl$)}]{\Gamma, x: A_1 \pl A_2 \vdash \lta{x} \casleft{x_1}{t_1} \casright{x_2}{t_2}: B; }
 {\Gamma, x_1:A_1 \vdash t_1:B & \Gamma, x_2:A_2 \vdash t_2:B}}
  & \\
&{\infer[\mbox{(right $\pl_1$)}]{\Gamma \vdash \lef t: A \pl B }
  {\Gamma \vdash t:A }}&{\infer[\mbox{(right $\pl_2$)}]{\Gamma \vdash \righ t: A \pl B }
  {\Gamma \vdash t:B }} \\
\end{tabular}
}
\end{center}
  \caption{Derived rules}\label{derivedrules}
\vspace*{-3mm}
\end{figure}

 The derived reduction rules for these constructions are:

$$\lt{t_1\te t_2}{x_1 \te x_2}{u}  \red  u[t_1/x_1, t_2/x_2]$$
$$\begin{array}{llll}
  \lta{\lef u} &\casleft{x_1}{t_1} && \\
               &\casright{x_2}{t_2} & \red & t_1\sub{u}{x_1} \\
 \lta{\righ u} &\casleft{x_1}{t_1} &&\\
               &  \casright{x_2}{t_2} & \red & t_2\sub{u}{x_2} 
\end{array}
$$
We also use as syntactic sugar, for $x$ a variable: $\lt{u}{x}{t}\stackrel{def}{=} ((\la x. t) \; u)$.

We now enlarge the language of types with a fix-point construction:
$$T::= \al \; |\; T \fm T \; |\; \forall \al. T \; |\; \bs \; T \;| \;
\mu \al. T$$
We add the corresponding typing rule and denote by ISALF,
intuitionistic light affine logic with fix-points, the new system:
Figure \ref{ISALFtyping}. If a pseudo-term is typable in ISALF then
clearly it is a well-formed term since these new rules do not have any
computational counterpart.

\begin{figure}[ht]
  \begin{center}
\fbox{
\begin{tabular}{l@{}ll}
  & the typing rules of ISAL and & \\
&& \\
  &{\infer[\mbox{(left unfold)}]{ x: A[\mu X.A/X], \Gamma \vdash t: B }
 {x: \mu X.A, \Gamma \vdash t: B}}
  &
{\infer[\mbox{(right unfold)}]{\Gamma \vdash t:A[\mu X.A/X]} 
  {\Gamma \vdash t:\mu X.A}}\\
&&\\
&{\infer[\mbox{(left fold)}]{ x: \mu X.A, \Gamma \vdash t: B }
 {x: A[\mu X.A/X], \Gamma \vdash t: B}}
&
 {\infer[\mbox{(right fold)}]{\Gamma \vdash t:\mu X.A} 
  {\Gamma \vdash t:A[\mu X.A/X]}} \\
\end{tabular}
}
\end{center}
  \caption{ISALF typing rules}\label{ISALFtyping}
\vspace*{-3mm}
\end{figure}
\begin{prop}[Subject reduction]
If we have in the system ISALF $\Gamma \vdash t:A$ and $t \red t'$
then $\Gamma \vdash t':A$.
\end{prop}    
Basically this result follows from the fact that as a logical system
ISALF admits cut-elimination.

Note that even though we have no restriction on the types on which we take fix points,
the typed terms are always normalizable and have a polynomial bound on
the length of their reduction. This follows from the fact that the polynomial termination
result (Theorem \ref{theopolytime}) already holds for untyped terms.

In the following we will handle terms typed in ISALF. Rather than
giving the explicit type derivations in the previous system, which is
a bit tedious because it is a sequent-calculus style presentation, we
will use a Church typing notation. The recursive typing rules and
second-order rules will be left implicit. From this notation it is
possible to reconstruct an explicit type derivation if needed.

Here is an example of typed term (integer 2 in unary representation)
 $$ \la s^{\bs (\al \fm \al)}. \la x^{\al}. \lt{s}{\bs s'}{(s' \; (s' \;x))^{\al}} \; : N$$
\section{Datatypes and list processing}\label{datatypes}

\subsection{Datatypes for lists}
Given a type $A$, we consider the following types defining lists of elements of $A$:
\begin{eqnarray*}
\ma{L}(A) &=& \forall \alpha. \bs(A \fm \alpha \fm \alpha)\fm \alpha \fm \alpha\\
L(A)     &=& \mu X.(1 \pl (A \te X))\\
\end{eqnarray*}

The type $\ma{L}(A)$ is the adaptation of the usual system F type for
lists. It supports an iteration scheme, but does not enable to define
in soft lambda-calculus a $cons$ function with type $\ma{L}(A) \fm A
\fm \ma{L}(A)$. This is analog to the fact that $\Nat$ does not allow
a successor function with type $\Nat \fm \Nat$ (\cite{Lafont02}).

 The type $L(A)$ on the contrary allows to define the usual elementary
 functions on lists $cons$, $tail$, $head$, but does not support
 iteration.


 The empty list for type $L(A)$ is given by
$\epsilon=\lef 1$ and the elementary functions by:
 $$\begin{array}{lll}
 cons &:& L(A) \fm A \fm L(A)  \\
 cons &=& \la l^{L(A)}. \la a^A . \righ {(a \te l)}\\
 tail &:& L(A) \fm L(A)  \\
 tail &=& \la l^{L(A)}. \lta{l} \casleft{l'}{\lef l' } \\
      &&  \qquad \qquad \qquad                         \casright{l'}{} \\
      &&   \qquad \qquad \qquad \qquad                {\lt{l'}{a \te l''}{l''}}\\
  head &:& L(A) \fm A  \\
   head &=& \la l^{L(A)}. \lta{l} \casright{l'}{} \\
        && \qquad \qquad \lt{l'}{a \te l''}{a}
\end{array}
$$
We would like to somehow bring together the advantages of
$\ma{L}(A)$ and $L(A)$ in a single datatype. This is what we will try
to do in the next sections.
\subsection{Types with integer}
Our idea is given a datatype $A$ to add to it a type $N$ so as to be
able to iterate on $A$.  The type $N \te A$ would be a natural
candidate, but it does not allow a suitable iteration.  We therefore
consider the following type:
 $$N[A] =  \forall \alpha. \bs(\alpha \fm \alpha)\fm \alpha \fm (A \te \alpha) $$

 Given $n$ integer and $a$ closed term of type $A$, we define an element of $N[A]$:
$$ n[a]= \la s^{\bs (\al \fm \al)}. \la x^{\al}. a^{A} \te \lt{s}{\bs
s'}{(s' \; s' \dots \; s' x)^{\al}} \; : N[A]$$
where $s'$ is repeated $n$ times.

We can give terms allowing to extract from an element $n[a]$ of type $N[A]$ either the data $a$
or the integer $n$.
 $$\begin{array}{lccc}
  extractd:         & N[A] &\fm& A\\
  extractint:       & N[A] &\fm& N \\
 \end{array}
$$
For instance 

$extractd=\la p^{N[A]}. \lt{(p \; \bs id^{\beta \fm \beta} \; id^{\al \fm \al} )}{a^A\te r^{\al}}{a}$

where $id$ is the identity term and $\beta= \al \fm \al$.

 However it is (apparently) not possible
 to extract both the data and the integer with a term of type $ N[A] \fm N \te A$. On the contrary from 
$n$ and $a$ one can build $n[a]$ of type $N[A]$:

$\begin{array}{lccc}
  build &:        & N \te A  \fm  N[A] \\
  build &=& \la t. \lt{t}{n\te a}{\la s.\la x. (n \; s \; x) \te a}
\end{array}
$

We can turn the construction $N[.]$ into a functor: we define the action of $N[.]$ on a
 closed term $f: A \fm B$ by

$\begin{array}{ll}
 N[f]= \la p^{N[A]}. \la s^{\bs (\al \fm \al)}. \la x^{\al}. &\lt{(p \; s \; x)^{A \te \al}}{a \te r}{} \\
             & \quad (f\; a)^{B} \te r^{\al}
\end{array} $

Then $N[f]: N[A] \fm N[B]$, and $N[.]$ is a functor.

We have the following principles:

$$\begin{array}{lccc}
absorb:   & N[A] \te B & \fm & N[A \te B]\\
 out:          & N[A \fm B] &\fm & (A \fm N[B])\\
 \end{array}
$$

The term $absorb$ for instance is defined by:

$\begin{array}{llll}
absorb & =& \la t^{N[A] \te B}. \la s^{\bs(\al \fm \al)}. \la x^{\al}. \\
             &  & \quad \lt{t}{p \te b}{}  \\
             &  &  \qquad \lt{(p\; s\; x)}{a \te r}{} \\
             &  &  \qquad \qquad  (a \te b \te r)^{A \te B \te \al} 
\end{array}
$



\subsection{Application to lists}

In the following we will focus our interest on lists. We will use as a shorthand 
notation $L'(A)$ for $N[L(A)]$. The terms described in the previous section can
be applied in this particular case.

In practice here we will use the type $L'(A)$ with the following
meaning: the elements $n[l]$ of $L'(A)$ handled are expected to be
such that the list $l$ has a length inferior or equal to $n$. We will then be able
 to do iterations on a list up to the length of the list.
 
%
%

 The function $erase$ maps $n[l]$ to $n[\epsilon]$ where $\epsilon$ is
 the empty list; it is obtained by a small modification on $exint$:
 \begin{eqnarray*}
erase &:& L'(A) \fm L'(A) \\
erase &=& \la p^{L'(A)}. \la s^{\bs(\al \fm \al)}.\la x^{\al}.\lt{(p\; s\; x)}{l^{L(A)}\te r^{\al}}{\epsilon^{L(A)} \te r^{\al}}
\end{eqnarray*}

We have for the type $L'(A)$ an iterator given by:
 \begin{eqnarray*}
Iter &:& \forall \alpha. \bs (\alpha \fm \alpha) \fm \alpha \fm L'(A) \fm (L(A) \te \alpha) \\
Iter &=& \la F^{\bs (\alpha \fm \alpha) }. \la e^{\alpha}. \la l^{ L'(A)}. (l \; F \; e)
\end{eqnarray*}

 If $F$ has type $B\fm B$, $e$ type $B$ and $F$ has free variables
 $\vect{x}$ then if $f=(Iter \; (\lt{\vect{y} }{\bs \vect{x} }{\bs
 F})\; e)$ we have:
 $$(f\; n[l]) \red l \te (\lt{\vect{y} }{\bs \vect{x} }{(F \dots (F \;
 e)\dots)},$$ where in the r.h.s. term $F$ is repeated $n$ times. Such
 an iterator can be in fact described more generally for any type
 $N[A]$ instead of $N[L(A)]$.

 Using iteration we can for instance build a function which reconstructs
 an element of $L'(A)$; it acts as an identity function on $L'(A)$ but is
 interesting though because in the sequel we will need to
 \emph{consume and restore} integers in this way:
\begin{eqnarray*}
reconstr &:& L'(A) \fm L'(A) \\
 F &:& \bs(\al \fm \al) \mbox{ with } \FV{F}=\{s^{\bs(\al \fm \al)}\} \\
 F &=& \lt{s}{\bs s'^{\al \fm \al}}{\bs(\la r^{\al}. (s' r)^{\al})}\\
reconstr &=& \la p^{L'(A)}. \la s^{\bs(\al \fm \al)}.\la x^{\al}.( Iter \; F\; x\; p)
\end{eqnarray*}

Given terms $t: A \fm B$ and $u: B \fm C$ we will denote by $t;u: A \fm C$ the \textit{composition}
of $t$ and $u$ defined as $(\la a^A. (u \; (t\; a)))$.
 
Finally we have the usual functions on lists with type $L'(A)$, using the ones defined before for the type $L(A)$:
$$\begin{array}{lllll}
 tail' &=& N[tail] &:& L'(A) \fm L'(A)  \\
 head' &=& N[head]; extractd &:& L'(A) \fm A  \\
cons' &=& N[cons]; out &: &L'(A) \fm A \fm L'(A)  
 \end{array}
 $$
 Note that to preserve the invariant on elements of $L'(A)$
 mentioned at the beginning of the section we will need to apply
 $cons'$ to elements $n[l]$ such that $n \geq m+1$ where $m$ is the
 length of $l$.
\subsection{Example: insertion sort}
We illustrate the use of the type $N[L(A)]$  by giving
the example of the insertion sort algorithm. Contrarily to the setting of Light affine logic
with system F like types, we can here define functions obtained by successive nested structural recursions.
Insertion sort provides such an example with two recursions. We use the presentation of this algorithm 
described in \cite{Hofmann00}.  

 The type $A$ represents a totally ordered set (we denote the order by
 $\leq$). Let us assume that we have for $A$ a comparison function
 which returns its inputs:
$$ comp: A \te A \fm A \te A, \; \mbox{ with } (comp\;  a_0 \; a_1) \red
 \left\lbrace\begin{array}{ll} a_0 \te a_1 & \mbox{ if } a_0 \leq a_1 \\
                              a_1 \te a_0  & \mbox{ otherwise }
              \end{array}\right.
$$
\textbf{Insertion in a sorted list.}

 Let $a_0$ be an arbitrary element of type $A$.
We will do an iteration on type: $B=L(A) \fm A \fm L(A) \te \al$. The iterated function will 
 reconstruct the integer used for its iteration. Let us take $F: \bs (B \fm B)$ with
 $\FV{F}=\{s^{\bs(\al \fm \al)}\}$, given by:

$\begin{array}{llrll}
F & =& \lt{s}{\bs s'^{\al \fm \al}}{} &&\\
  &  &  \qquad \bs (\la \phi^B. \la l^{L(A)}. \la a^A. &&\\
  &  & \qquad \lta{l} & &\\
  &&   &\casleft{l_1}{\lt{(\phi \; \epsilon \; a_0)}{l' \te r^{\al}}{}}     \qquad &\backslash \star  \mbox{ case l empty } \\
  &  & &   \qquad \qquad (cons \; a \; \epsilon)^{L(A)} \te (s'\; r)^{\al} &\\
 &   & &  \casright{l_1}{\lt{l_1}{b \te l'}{} } \qquad &\backslash \star  \mbox{ case l non empty }\\
 & &   &\qquad \qquad \lt{(comp \; a \;b)}{a_1 \te a_2}{} &\\
 & &   &\qquad \qquad \qquad \lt{(\phi\; l'\; a_2)}{l'' \te r}{} &\\
  & &  & \qquad \qquad \qquad \qquad (cons \; a_1 \; l'')\te (s'\; r)^{\al}&
\end{array}
$

Let $e:B$ be the term $e=\la l^{L(A)}. \la a^{A}. ({\epsilon}^{L(A)} \te x^{\al})$. Note that $\FV{e}=\{x^{\al}\}$. Then we have:
$$ s: \bs{\al \fm \al}, x:\al \vdash (Iter \; F \; e): L'(A) \fm L(A)\te B$$
Finally we define:

$\begin{array}{llrl}
insert & =& \la p^{L'(A)}. \la a^A. &\la s^{\bs(\al \fm \al)}. \la x^{\al}\\
       &&                           & \lt{(Iter \; F \; e\; p)^{L(A)\te B}}{(l^{L(A)}\te f^B)}{}\\
       &&&                             \qquad \qquad (f\; l\; a)^{L(A) \te \al}
\end{array}
$

and get: $insert: L'(A) \fm A \fm L'(A)$.

\textbf{Insertion sort.}

 We define our sorting program by iteration on $B=L(A) \te L'(A)$. The
left-hand-side list is the list to process while the right-hand-side
one is the resulting sorted list. Then $F: \bs (B \fm B)$ is the
closed term given by:

$\begin{array}{llll}
F & =&\bs (\la t^B. \lt{t}{l_1^{L(A)} \te p^{L'(A)}}{\lta{l_1}}& \\
  && \qquad    \qquad           \casleft{l_2} (\lef l_2) \te p             & \backslash \star \mbox{ case $l_1$ empty }\\
&&   \qquad    \qquad           \casright{l_2} {\lt{l_2}{a \te l_3}{}}             & \backslash \star \mbox{ case $l_1$ non empty }\\
&&                  \qquad \qquad \qquad  l_3^{L(A)} \te (insert \; p \; a)^{L'(A)} & 
\end{array}
$

$e=l^{L(A)} \te (erase \; p_0)^{L'(A)}:B$

We then have:

$l:L(A), p_0: L'(A) \vdash (Iter \; F \; e): L'(A) \fm L(A) \te B $

So we define:

$\begin{array}{lll}
presort & =&\la p_0^{L'(A)}. \la p_1^{L'(A)}. \la p_2^{L'(A)}.\\
        &  & \qquad \lt{(exlist \; p_1)}{l^{L(A)}}{}\\
        &  & \qquad \qquad \lt{(Iter \; F \; e \; p_2)}{l' \te l'' \te p'}{l''}
\end{array}
$

Using multiplexing we then get:

$\begin{array}{lll}
sort & =&\la p^{\bs L'(A)}. \lt{p}{\bs p'^{L'(A)}}{}\\
     && \qquad (presort \; p' \; p' \; p' )^{L'(A)}
 \end{array}
$

So:

 $sort: \bs L'(A) \fm L'(A)$.

\begin{remark}
More generally the construction $N[.]$ can be applied successively to
define the following family of types:
\begin{eqnarray*}
N^{(0)}[A]      &=& A \\
N^{(i+1)}[A]      &=&  N[N^{(i)}[A]]  
\end{eqnarray*}
 This allows to type programs obtained by several nested structural
recursions. For instance insertion sort could be programmed with type
$N^{(2)}[A] \fm N^{(2)}[A]$. This will be detailed in a future work.
\end{remark}
\subsection{Iteration}

 We saw that with the previous iterator $Iter$ one could define from
 $F:B\fm B$ and $e: B$ an $f$ such that: $(f\; l[n]) \red l \te
 (\lt{\vect{y} }{\bs \vect{x} }{(F \dots (F \; e)\dots)}$. However the
 drawback here is that $l$ is not used in $e$. We can define a new
 iterator which does not have this default, using the technique
 already illustrated by the \emph{insertion} term.  Given a type
 variable $\al$, we define $C=L(A) \fm \al$.

 If $g$ is a variable of type $\bs (\al \fm \al)$, we define:

$$ G'= \lt{g}{\bs g'}{\bs(\la b'^{C}. \la l^{L(A)}. (g' \; (b' \; l)))}: \bs (C \fm C)$$ 
Then:
$$\begin{array}{lll}
It & =&\wedge \al . \la g^{\bs ({\al}\fm {\al})}. \la e^{C}. \la p^{L'(A)}. \\
        &  & \qquad \qquad \lt{(Iter \; G' \; e^{C} \; p)}{l_1^{L(A)}\te f^{C} }{} \\
        &  & \qquad \qquad \qquad  \qquad \qquad (f \; l_1)\\
It & : & \forall \al.  \bs (\alpha \fm \alpha) \fm (L(A) \fm \alpha) \fm L'(A) \fm \alpha \\
\end{array}
$$

Then if $f=(It \; (\lt{\vect{y} }{\bs \vect{x} }{\bs F})\; \la l_0.e')$ we have:
$$(f\; l[n]) \red \lt{\vect{y} }{\bs \vect{x} }{(F \dots (F \; e'\sub{l}{l_0} )\dots)},$$
where in the r.h.s. term $F$ is repeated $n$ times.

In appendix \ref{map} we give an example of use of this new iterator to program a map function.
\section{Conclusion and future work}\label{conclusion}

We studied a variant of lambda-calculus which can be typed in Soft Affine
Logic and is intrinsically polynomial. The contribution of the paper is twofold:
\begin{itemize}
\item We showed that the ideas at work in Soft Linear Logic to control
  duplication can be used in a lambda-calculus setting with a concise
  language. Note that the language of our calculus is  simpler than
  those of calculi corresponding to ordinary linear logic such as in \cite{BBPH93}, \cite{Abramsky93}. Even 
  if the underlying intuitions come from proof-nets and Lafont's results, 
  we think that this new presentation will facilitate further study of 
  Soft logic.

\item We investigated the use of recursive types in conjunction with Soft
  logic. They allowed us to define non-standard types for lists and we
  illustrated the expressivity of Soft lambda-calculus by programming the
  insertion sort algorithm.
\end{itemize}
We think Soft lambda-calculus provides a good framework to study the
  algorithmic possibilities offered by the ideas of Soft logic. One 
  drawback of the examples we gave here is that their programming is
  somehow too low-level. One would like to have some generic way of
  programming functions defined by structural recursion (with some
  conditions) that could be compiled into Soft lambda-calculus. Current 
  work in this direction is under way with
  Kazushige Terui. It would be interesting to be able to state sufficient conditions
 on algorithms, maybe related to space usage, for being
  programmable in Soft lambda-calculus.

\bibliography{soft}

\newpage

\appendix
\begin{center}{\Large APPENDIX}\end{center}
\section{Some proofs of section \ref{softlambda-calculus} }

\subsection{Lemma \ref{substitution1}}
\begin{proof}
  We proceed by induction on $t$.
\begin{itemize}
\item  The cases where $t$ is a variable or an abstraction
 are straightforward. 
\item  If $t= \bs t_1$ then $\FV{t}=\TV{t}$, so as $x \notin \TV{t}$ then
 $x \notin \FV{t}$. Therefore $t \sub{u}{x} =t$ and the result follows.

\item If  $t=\lt{t_1}{\bs y}{t_2}$ then we have:

$t\sub{u}{x}=\lt{t_1\sub{u}{x} }{\bs y}{t_2\sub{u}{x} } $.

As $\TV{t_1} \subseteq \TV{t}$ we know that $x \notin \TV{t_1}$ and
$t_1$, $u$ satisfy the hypothesis of the statement, so by induction
hypothesis on $t_1$ we have that $t_1\sub{u}{x}$ is a term and
 $\TV{t_1\sub{u}{x} }= \TV{t_1}$. Similarly $t_2\sub{u}{x}$ is a term and
 $\TV{t_2\sub{u}{x} }= \TV{t_2}$.

So we have:
 \begin{eqnarray}
 \TV{t_1\sub{u}{x} }&=& \TV{t_1}\\
\FV{t_2\sub{u}{x} } &\subseteq& \FV{t_2 } \cup \FV{u} \backslash \{x\}, \\
\TV{t_1} \cap \FV{t_2} &=& \emptyset  \mbox{ (because $t$ is a term)}\\
\TV{t_1} &\subseteq& \TV{t}\\
\TV{t} \cap \FV{u} &=& \emptyset \mbox{ (by assumption) }
\end{eqnarray}
From (4) and (5) we get: $\TV{t_1} \cap \FV{u} = \emptyset$. From this result and (3), (2) we deduce:
$ \TV{t_1 } \cap \FV{t_2\sub{u}{x} }= \emptyset$.
So, with (1):
 $ \TV{t_1\sub{u}{x} } \cap \FV{t_2\sub{u}{x} }= \emptyset$.

In the same way one can check that $ \TV{t_2\sub{u}{x} } \cap \FV{t_1\sub{u}{x} }= \emptyset$. It follows 
that $t\sub{u}{x} =\lt{t_1\sub{u}{x} }{\bs y}{t_2\sub{u}{x} }$ is a term and:
\begin{eqnarray*}
\TV{t\sub{u}{x} }&=& \TV{t_1\sub{u}{x} } \cup \TV{t_2\sub{u}{x} }\backslash\{y\}\\
                 &=& \TV{t_1} \cup \TV{t_2} \backslash\{y\}\\
                 &=& \TV{\lt{t_1}{\bs y}{t_2}}\\
                 &=& \TV{t}
\end{eqnarray*}

\item The case $t=(t_1 \; t_2)$ is handled in a similar way as the previous one.
\end{itemize}
\end{proof}

\subsection{Lemma \ref{substitution1bis}}
\begin{proof}
  The proof is by induction on $t$.
\begin{itemize}
\item Again the cases where $t$ is a variable or an abstraction
 are straightforward.
\item  If $t= \bs t_1$ then
  the hypothesis of the statement cannot be met as we have
  $\FV{t}\backslash \TV{t}=\emptyset$.
\item The cases $t= \lt{t_1}{\bs y}{t_2}$ or $t=(t_1 \, t_2)$ 
 are quite similar, so let us just handle
  one of them, for instance this time $t=(t_1 \, t_2)$.
  
  As $\no{x}{t}=1$ we have: either  $\no{x}{t_1}=1$ and $\no{x}{t_2}=0$, or
  the converse. Let us assume for instance $\no{x}{t_1}=1$ and
  $\no{x}{t_2}=0$. Then as $\FV{t_1} \subseteq \FV{t}$ and $\TV{t_1}
  \subseteq \TV{t}$ we know that $t_1$, $u$  satisfy the conditions. By
  induction hypothesis on $t_1$ we deduce that $t_1\sub{u}{x} $ is a
  term and $\TV{t_1\sub{u}{x} }= \TV{t_1} \cup \TV{u} $. Besides,
  $\FV{t_1\sub{u}{x} }= \FV{t_1} \cup \FV{u} $.
  
  So we have $\TV{t_2} \cap \FV{t_1 \sub{u}{x} }= \emptyset$ and

$\FV{t_2} \cap \TV{t_1 \sub{u}{x} }=\FV{t_2} \cap (\TV{t_1} \cup \TV{u})=\emptyset$.

So $(t_1 \sub{u}{x} t_2)$ is a term, and:
$$\TV{t_1 \sub{u}{x} \; t_2}= \TV{t_1} \cup \TV{u} \cup \TV{t_2}= \TV{t} \cup \TV{u}.$$
\end{itemize}
\end{proof}

\subsection{Lemma \ref{substitution2}}
\begin{proof}
  We proceed by induction on $t$.
\begin{itemize}
\item if $t$ is a variable then $\TV{t}=\emptyset$, which contradicts the assumption that $x \in \TV{t}$.
\item if $t=\la y. t_1$, then $x \in \TV{t_1}$. By induction hypothesis
  on $t_1$, $t_1\sub{u}{x} $ is a term.  As $y \notin \FV{u}$ and
  $\no{y}{t}\leq 1$ we have $\no{y}{t\sub{u}{x} }\leq 1$, and so $\la y.
  t_1\sub{u}{x} $ is a term. Moreover:
\begin{eqnarray*}
\TV{\la y. t_1\sub{u}{x} }&=& \TV{t_1\sub{u}{x} }= \TV{t_1} \backslash\{x\} \cup \FV{u}\\
&=&\TV{t} \backslash\{x\} \cup \FV{u}
\end{eqnarray*}  

\item if $t= \bs t_1$, then $\TV{t_1}=\emptyset$. So $x \notin
  \TV{t_1}$ and $\TV{t_1} \cap \FV{u}=\emptyset$, and applying lemma
  \ref{substitution1} we get: $t_1\sub{u}{x} $ is a term and
  $\TV{t_1\sub{u}{x} }=\emptyset$.

Moreover $\FV{t_1\sub{u}{x} }=\FV{t_1} \cup \FV{u}$, and as $t_1$, $u$
are both well-formed and $\FV{u} \cap \FV{t_1}=\emptyset$ we get that
$t_1\sub{u}{x} $ is well-formed. It follows that $t_1\sub{u}{x} $ is a
term, that is to say that $t\sub{u}{x} $ is a term, and:

$\TV{t\sub{u}{x} }= \FV{t_1 \sub{u}{x} }= \FV{t_1}\backslash\{x\}\cup
\FV{u}=\TV{t}\backslash \{x\}\cup \FV{u}.$
\item if $t=(t_1 \; t_2)$  then either $x\in \TV{t_1}$ and $x \notin \TV{t_2}$, or
 $x\notin \TV{t_1}$ and $x \in \TV{t_2}$. Let us assume for instance $x\in \TV{t_1}$ and
 $x \notin \TV{t_2}$. We have $\FV{t_i} \cap \FV{u}=\emptyset $ for $i=1, 2$. By induction
hypothesis on $t_1$ we have $t_1\sub{u}{x} $ is a term and
 $\TV{t_1\sub{u}{x} }=\TV{t_1} \backslash \{x\}\cup \FV{u}$. Moreover as $t_2\sub{u}{x} =t_2$,
$t_2\sub{u}{x} $ is also a term. We have:

$\FV{t_1\sub{u}{x} }=\FV{t_1}\backslash \{x\}\cup \FV{u}$, so $\TV{t_2}\cap \FV{t_1\sub{u}{x} }=\emptyset$,

$\TV{t_1\sub{u}{x} }=\TV{t_1} \backslash \{x\}\cup \FV{u}$, so $\FV{t_2}\cap \TV{t_1\sub{u}{x} }=\emptyset$.

So $(t_1\sub{u}{x} \; t_2)$ is a term, that is to say $t\sub{u}{x} $ is a term, and

$\TV{t\sub{u}{x} }=\TV{t_1\sub{u}{x} }\cup\TV{t_2}= \TV{t}\backslash\{x\} \cup \FV{u}$.

\item the case $t=\lt{t_1}{\bs y}{t_2}$ is handled in a similar way.
\end{itemize}
\end{proof}







\section{Proof of lemma \ref{lemKey}}\label{prooflemKey}
\begin{proof} 
\begin{enumerate}
\item proof by induction on $t$ considering $x\in\FV{t}$ or not.
\item by induction on $t$ we have: 
\begin{itemize}
\item if $t=\la y.t_1$ then $x\in\TV{t_1}$. By induction hypothesis we have 
$\WE{t\sub{u}{x} }{n}\leqslant\WE{t}{n} +n\WE{u}{n}$.
\item if $t=\bs t_1$ then by definition of terms 
  $x\in\TV{t} =\FV{t_1}$, $\TV{t_1} =\emptyset$ and $\no{x}{t} =1=\no{x}{t_1}$. 
The result holds.
\item if $t=(t_1 t_2)$ then either $x\in\TV{t_1}$ and
  $x\notin\FV{t_2}$ or $x\in\TV{t_2}$ and $x\notin\FV{t_1}$. 
In the first case 
$\WE{t\sub{u}{x} }{n} =\WE{t_1\sub{u}{x} \; t_2\sub{u}{x} }{n} =
\WE{t_1\sub{u}{x} }{n} +\WE{t_2\sub{u}{x} }{n}
\leqslant\WE{t_1}{n} +n\WE{u}{n} +\WE{t_2}{n}
\leqslant\WE{t_1t_2}{n} +n\WE{u}{n}$.
The second case is similar.
\item
 if $t=\lt{u}{\bs x}{t_1}$ then because there is the following disjoint union 
$\TV{t}= \TV{u}\uplus (\TV{t_1}\setminus\{ x\} )$, the result holds.
\end{itemize}
\end{enumerate}
\end{proof}

\section{Example: map function}\label{map}

We use the iterator $It$ to define the map function. Let $B= L(A) \te L(C) \te \al$. We consider variables $f'^{A\fm C}$ and $s'^{\al \fm \al}$.
$$\begin{array}{llll}
F & =& \la t^{B}. \lt{t}{l_1^{L(A)} \te l_2^{L(C)} \te r^{\al}}{} &\\
 &  &  \qquad \lta{l_1} &\\
 &&   &\casleft{l'_1}{(\lef l'_1) \te l_2 \te (s'\; r)} \\
 &&   &\casright{l'_1}{}   \\
 & &  &  \qquad (tail \; l'_1) \te (cons \; (f' \; (head \; l'_1)) \; l_2) \te (s' \; r)\\
F &:& B \fm B &\\
&&&\\
e&=& \la l_0. l_0^{L(A)} \te \epsilon \te x^{\al} &\\
e&: & L(A) \fm B & 
\end{array}
$$
We then define $\phi: L'(A) \fm B$ by:
$$
\begin{array}{lll}
 \phi&=&(It \; (\lt{f^{\bs (A \fm C)}}{\bs f'}{}\\
     &&     \qquad \qquad                 \lt{s^{\bs (\al \fm \al)}}{\bs s'}{\bs F}) \; e^{L(A) \fm B})
\end{array}
$$
 We can then define a $map$ function, which however reverses the order of the elements of the list. To obtain the proper $map$ function we would have to compose it with a reverse function.

$$
\begin{array}{lll}
 map&=& \la f^{\bs(A\fm C)}.\la p^{L'(A)}. \la s^{\bs(\al \fm \al)}.\la x^{\al}. \\
    & & \qquad \lt{(\phi^{L'(A)\fm B} \; p)}{l_1 \te l_2 \te r}{}\\
    & & \qquad \qquad \qquad l_2 \te r\\
map &:& \bs (A \fm C) \fm L'(A) \fm L'(C)
\end{array}
$$
\end{document}